\documentclass[aps,prb,twocolumn,longbibliography,groupedaddress,showpacs,floatfix,superscriptaddress]{revtex4}

\usepackage{soul} % to use \st for strikeout 
\usepackage{epsfig}
\usepackage{amsmath,amssymb}
\usepackage{graphicx}
\usepackage[dvipsnames,usenames]{color}
\usepackage[normalem]{ulem}
\emergencystretch=\maxdimen
\begin{document}

\title{Compressible Ferrimagnetism in the depleted Periodic Anderson
Model}
\author{N.C. Costa} 
\affiliation{Instituto de F\'isica, Universidade Federal do Rio de Janeiro
Cx.P. 68.528, 21941-972 Rio de Janeiro RJ, Brazil}
\affiliation{Department of Physics, University of California, Davis, CA
95616, USA}
\author{M.V. Ara\'ujo} 
\affiliation{Departamento de F\'isica, Universidade Federal do Piau\'i,
Teresina PI, Brazil}
\author{J.P. Lima} 
\affiliation{Departamento de F\'isica, Universidade Federal do Piau\'i,
Teresina PI, Brazil}
\author{T. Paiva} 
\affiliation{Instituto de F\'isica, Universidade Federal do Rio de Janeiro
Cx.P. 68.528, 21941-972 Rio de Janeiro RJ, Brazil}
\author{R.R. dos Santos} 
\affiliation{Instituto de F\'isica, Universidade Federal do Rio de Janeiro
Cx.P. 68.528, 21941-972 Rio de Janeiro RJ, Brazil}
\author{R.T. Scalettar}
\affiliation{Department of Physics, University of California, Davis, CA
95616, USA}
\begin{abstract}
Tight-binding Hamiltonians with single and multiple orbitals exhibit an
intriguing array of magnetic phase transitions.  In most cases the
spin ordered phases are insulating, while the disordered phases
may be either metallic or insulating.  In this paper we report a Determinant
Quantum Monte Carlo study of interacting electrons in a geometry
which can be regarded as a two-dimensional Periodic Anderson Model
with depleted interacting ($f$) orbitals. For a single depletion, we observe
an enhancement of antiferromagnetic correlations and formation of localized states.
For half of the $f$-orbitals regularly depleted, the system exhibits a ferrimagnetic
ground state. We obtain a 
quantitative determination of the nature of magnetic order,
which we discuss in the context of Tsunetsugu's theorem,
and show that, although the dc conductivity indicates insulating
behavior at half-filling, the compressibility remains finite.
\end{abstract}

%\date{Version 6.3 -- \today}
\date{\today}

\maketitle

%%%%%%%%%%%%%%%%%%%%%%%%%%%%%%%%%%%%%%%%%%%%%%%%%%%%%%%%%%%%%%%%%%
\section{Introduction}
%%%%%%%%%%%%%%%%%%%%%%%%%%%%%%%%%%%%%%%%%%%%%%%%%%%%%%%%%%%%%%%%%%

Tight binding Hamiltonians provide insight into many of the properties
of strongly correlated electron systems, from magnetism and
metal-insulator transitions, to superconductivity and charge
ordering\cite{gebhard97,fazekas99}.  The simplest of these, the single
band Hubbard model (HM), is known, for example, to be insulating and to
exhibit long range antiferromagnetic (AF) order at half-filling on a
square lattice \cite{hirsch85,hirsch89} for any ratio of the on-site
interaction $U$ to hopping $t$, and to undergo a paramagnetic metal to
insulating AF transition above a nonzero critical $U_c$ on other
geometries such as the honeycomb lattice.  The Nagaoka
theorem\cite{nagaoka66} notwithstanding, the ferromagnetic behavior
which is robust within mean-field theory\cite{fazekas99,hirsch85}, seems
to be difficult to achieve when the single band Hamiltonian is solved
with more exact methods\cite{tasaki98}.

The generalization of tight-binding Hamiltonians to multiple bands opens
up a richer variety of magnetic behavior.  In the case of the periodic
Anderson model (PAM), the interplay of the on-site repulsion $U_f$ on
localized ($f$) orbitals with the hybridization $V$ to a noninteracting
conduction ($d$) band results in a competition of long range magnetic
order arising from the Ruderman-Kittel-Kasuya-Yosida (RKKY) interaction
(at small $V$) and spin liquid behavior (at large $V$).\cite{Vekic95,stewart01,DeLeo08,Hu17}  Not uncommonly
in these more complex situations, orbital ordering coexists with spin
ordering\cite{cameron16}.

Many of these models offer quite remarkable insights into strongly
correlated materials; for instance, the HM replicates several prominent
qualitative features of cuprate superconductors, such as
the AF and $d$-wave pairing, as well as stripe formation.\cite{scalapino94,Zhang97,Millis98,Maier00,Capone07,Wu15}
The AF-singlet transition and strongly renormalized effective electronic mass
in the PAM, and its strong coupling limit (the Kondo lattice model) helps
to explain different ground states in heavy-fermion
materials.\cite{stewart84,Hirsch86,Fye90,Fye91,Costi88,stewart01}  Manganites\cite{dagotto01} and
iron-pnictide superconductors\cite{si16} are also materials for which
appropriate multi-orbital models have been useful for developing
understanding of magnetism, pairing, charge order, and transport.

Geometries which can be regarded as arising from regular `depletions' of
the square lattice HM have also been explored, both to answer
fundamental questions about types of magnetic order and for the
understanding of specific materials.  An example of the former is Lieb's
theorem\cite{lieb89}, which rigorously demonstrates 
that a 1/4-depleted square lattice 
possesses ground state ferrimagnetic behavior.
Instances of the latter are the 1/5-depleted square lattice which can
explain spin liquid behavior in CaV$_4$O$_9$ \cite{troyer96,khatami14},
and the 1/3-depleted square lattice which shed light into the properties of
layered nickelates such as La$_{4}$Ni$_{3}$O$_{8}$\cite{Guo17,Botana16,Zhang16}.
In these situations, the depletion converts the initial single band
nature to a model with multiple bands.  In the case of the Lieb lattice,
one of these bands is dispersionless, a feature which is intimately tied
to the appearance of ferromagnetism.  (Random) site depletion of
tight-binding Hamiltonians has also been used to understand the effects
of the substitution of nonmagnetic atoms for magnetic ones, for example
the replacement of Cu by Zn in cuprate materials.\cite{Xiao88,Keimer92,Mahajan94,Julien00}  In cases where the
underlying geometry contains triangular lattice coordination, depletion
can aid AF behavior by reducing frustration,\cite{coles87} eg in CeAl$_3$.
More complex cases, such as depletion in the PAM seems relevant to understand
the formation of magnetism in heavy fermion materials.
In this case, mean-field and perturbation theory\cite{Potthoff14,Potthoff15_2,Seki16}
have provided evidence of a ferromagnetic ground state.

In view of this, we investigate the combination of these two avenues, a
PAM which {\it begins} already with two bands, but is then
subject to site depletion.  Our main conclusion is that depletion 
can drive the PAM into a magnetically ordered state, even for parameter choices
which are deep in the singlet phase for the undepleted lattice.  In addition to
obtaining its magnetic phase diagram quantitatively, we will show that
an unusual property develops in which the ordered regime is also
compressible.  

The organization of this paper is as follows:  In Sec.~II
we define the tight-binding Hamiltonian precisely, review the
Determinant Quantum Monte Carlo (DQMC) methodology\cite{blankenbecler81}
briefly, and define the observables used to characterize the model's
properties.  Section III presents data on the effect of the removal of
a single site; the resulting enhanced spin response provides an initial
clue to the robustness of magnetism in the regularly depleted geometry,
described in Sec.~IV.  Section V analyzes data for the compressibility,
effective hopping, and conductivity, and Sec.~VI contains our
conclusions.

%%%%%%%%%%%%%%%%%%%%%%%%%%%%%%%%%%%%%%%%%%%%%%%%%%%%%%%%%%%%%%%%%%
\section{Model} \label{Model and Method}
%%%%%%%%%%%%%%%%%%%%%%%%%%%%%%%%%%%%%%%%%%%%%%%%%%%%%%%%%%%%%%%%%%

The depleted PAM we consider here is described by the Hamiltonian
\begin{align} \label{eq:PAM_hamil}
\nonumber \hat {\mathcal H} = &
-t \sum_{\langle \mathbf{i}, \mathbf{j} \rangle, \sigma} 
\big(d^{\dagger}_{\mathbf{i} \sigma} d^{\phantom{\dagger}}_{\mathbf{j} \sigma} + {\rm h.c.} \big)
 - V \sideset{}{'}\sum_{ \mathbf{i} \sigma}
\big(d^{\dagger}_{\mathbf{i} \sigma} 
f^{\phantom{\dagger}}_{\mathbf{i} \sigma} + {\rm h.c.} \big) 
\\ \nonumber 
& + U_{f} \sideset{}{'}\sum_{\mathbf{i}}
\big(n^{f}_{\mathbf{i} \uparrow} - \frac{1}{2} \big) 
\big(n^{f}_{\mathbf{i} \downarrow} - \frac{1}{2} \big)
\\ \nonumber 
& + \sum_{\mathbf{i} \sigma} \epsilon^{d}_{\mathbf{i}}
d^{\dagger}_{\mathbf{i} \sigma} d^{\phantom{\dagger}}_{\mathbf{i} \sigma}
+ \sideset{}{'}\sum_{\mathbf{i} \sigma} \epsilon^{f}_{\mathbf{i}} 
f^{\dagger}_{\mathbf{i} \sigma} f^{\phantom{\dagger}}_{\mathbf{i} \sigma}
\\
& -\mu \sum_{\mathbf{i} \sigma}
d^{\dagger}_{\mathbf{i} \sigma} d^{\phantom{\dagger}}_{\mathbf{i} \sigma}
+ \sideset{}{'}\sum_{\mathbf{i} \sigma}
f^{\dagger}_{\mathbf{i} \sigma} f^{\phantom{\dagger}}_{\mathbf{i} \sigma}
,
\end{align}
where the unprimed sums over $\mathbf{i}$ run over a two dimensional
square lattice, with $\langle \mathbf{i}, \mathbf{j} \rangle$ denoting
nearest-neighbors, while the primed sums are restricted to the set of
sites having $f$-orbitals.  The specific depletion patterns will be
described in the coming sections.  The first term on the right-hand side of
Eq.\,\eqref{eq:PAM_hamil} represents the hopping of $d$-electrons, while
the second term contains the hybridization, $V$, between $d$ and
$f$-orbitals.  The Coulomb repulsion on localized $f$-orbitals is
included in the third term, with $n^{f}_{\mathbf{i} \sigma} =
f^{\dagger}_{\mathbf{i} \sigma} f^{\phantom{\dagger}}_{\mathbf{i} \sigma}$ being the number
operator of $f$-electrons.  The last two terms correspond to onsite
energies $\epsilon^{d}_{\mathbf{i}}$ and $\epsilon^{f}_{\mathbf{i}}$ of
$d$ and $f$-orbitals, respectively.  The hopping integral $t\equiv 1$
defines the scale of energy.

We analyze Eq.\,\eqref{eq:PAM_hamil} using DQMC, a numerically exact
technique in which all sources of error, statistical (from finite
sampling times) and systematic (from the discretization of the inverse
temperature $\beta$) can be removed to the desired degree of accuracy.
The basic idea of the method is the use of the Trotter-Suzuki
decomposition to separate the exponentials of the one-body and
two-body pieces, $\hat {\mathcal K}$ and $\hat {\mathcal P}$
respectively, in the partition function,
${\cal Z} = {\rm Tr}\, e^{-\beta \hat {\mathcal H} }
= {\rm Tr}\, \big[ \big(
e^{-\Delta\tau ( \hat {\mathcal K} + \hat {\mathcal P})} \big)^{l} \big] 
\approx {\rm Tr}\, \big[ 
e^{-\Delta\tau \hat {\mathcal K}} 
e^{-\Delta\tau \hat {\mathcal P}} 
e^{-\Delta\tau \hat {\mathcal K}} 
e^{-\Delta\tau \hat {\mathcal P}} \cdots \big]$.
Here $l=\beta/ \Delta \tau$ is the number of incremental
time evolution operators.
This decomposition has an error proportional to $(\Delta \tau)^2$ and is
exact in the limit $\Delta \tau \to 0 $.  The resulting isolation of
$e^{-\Delta \tau \hat {\mathcal P}}$ allows for the performance of a
discrete Hubbard-Stratonovich (HS) transformation so that it can be
rewritten in quadratic (single body) form, but with the cost of
introducing a discrete auxiliary field with components on each of the
space and imaginary time lattice coordinates.  The fermions are then
integrated out, and the HS field is sampled by the Monte Carlo
technique.  In the work we report here we choose $t \Delta \tau=0.125$
so that the error from the Trotter-Suzuki decomposition is less than, or
comparable to, that from the Monte Carlo sampling.  We therefore report
error bars from the latter.  More details about the method are discussed
in Ref.\,\onlinecite{Santos03} and references therein.

Although DQMC is exact, its low temperature application is restricted to
systems with particle-hole or other symmetries\cite{li16}, owing to the
minus-sign problem\cite{loh90,troyer05}.  For this reason, our focus is
on half-filling, $\mu = \epsilon^{f} = \epsilon^{d} = 0$, where the sign
problem is absent.  Fortunately, this density is of considerable
interest, both because of the strong magnetic order favored by
commensurate filling, and by the materials for which half-filling is
appropriate (e.g.~the undoped parent compounds of the cuprate
superconductors).  Depleting $f$-orbitals, i.e.  removing them from the
lattice, preserves particle-hole symmetry (PHS).  This is true
regardless of the number or pattern of the removed sites, in much the
same way that PHS is present for arbitrary (including
position-dependent) choices of the energy scales $t, V$, and $U_f$, as
long as the hopping only connects sites on opposite sublattices of a
bipartite lattice.

We concentrate on the following observables:
The magnetic features of the Hamiltonian will be characterized by the real
space spin-spin correlation function,
\begin{align}
C^{\alpha \gamma}({\bf j}) = 
\langle 
S^{\alpha,\, -}_{{\bf j_0}+{\bf j}} 
S^{\gamma,\, +}_{{\bf j_0}} 
\rangle 
=
\langle 
c^{\alpha \,\,\dagger}_{{\bf j_0}+{\bf j} \, \downarrow} 
c^{\alpha \phantom{\dagger}}_{{\bf j_0}+{\bf j} \, \uparrow} 
c^{\gamma \,\,\dagger}_{{\bf j_0} \, \uparrow} 
c^{\gamma \phantom{\dagger}}_{{\bf j_0} \, \downarrow} 
\rangle \,\,,
\label{eq:Cspin}
\end{align}
where the orbital indices are $\alpha, \gamma = d,f$.  (Later in the paper, 
we will use an alternate notation which further distinguishes
the two types of conduction electron orbitals, those with a partner $f$
orbital and those for which the partner has been removed.)  As the notation
suggests, $C^{\alpha \gamma}({\bf j}) $ is independent of ${\bf j_0}$ 
for translationally invariant geometries.  The Fourier transform of
$C^{\alpha \gamma}({\bf j})$ is the magnetic structure factor, 
\begin{align} 
S^{\alpha \gamma}({\bf q}) = 
%% \frac{1}{N} \sum_{\bf j} c^{\alpha \gamma}({\bf j})
\sum_{\bf j} C^{\alpha \gamma}({\bf j})
\, e^{i {\bf q} \cdot {\bf j}}. 
\label{eq:Sq} 
\end{align} 
%% with $N$ being the number of sites.

In addition to these equal time correlation functions, we also measure
appropriate unequal time quantities including the
magnetic susceptibility, 
\begin{align}
%% \chi^{\alpha\gamma}({\bf q}) = \frac{1}{N} \sum_{\mathbf{i} \mathbf{j}} 
\chi^{\alpha\gamma}({\bf q}) =  \sum_{\mathbf{j}} 
\int_{0}^{\beta} d\tau \langle 
S^{\alpha,\, -}_{{\bf j_0}+{\bf j}}(\tau)
S^{\gamma,\, +}_{{\bf j_0}}(0)
%% S^{z}_{\mathbf{i}}(\tau) S^{z}_{\mathbf{j}}(0) 
\rangle
\, e^{i {\bf q} \cdot {\bf j}} \, ,
\label{eq:hom.chi}
\end{align}
For $\chi$ we will mostly examine the uniform case, ${\bf q}=0$.
Although we have defined both the equal time correlations,
Eq.\,\eqref{eq:Cspin}, and susceptibility, Eq.\,\eqref{eq:hom.chi}, in terms
of the $xy$ ($+-$) spin components, these are, by symmetry, equivalent to those
in the $z$ direction.

Metal-insulator transitions are characterized via the electronic
compressibility ($\kappa$) and the dc conductivity ($\sigma_{dc}$).  The
former is defined as
\begin{equation}
\kappa = -\frac{1}{\rho^2}\frac{\partial \rho}{\partial \mu},
\label{eq:kappa}
\end{equation}
where $\rho$ is the electronic density.  Like the magnetic
susceptibility, which can be measured either by the response of the
magnetization to a small field or as a sum over spin correlation
functions as in Eq.\,\eqref{eq:hom.chi}, the compressibility can be
obtained via evaluating the change in particle density induced by a
shift in chemical potential via Eq.\,\eqref{eq:kappa} or by summing density
correlation functions.  We use the former approach here.  

The conductivity, $\sigma_{dc}$, is evaluated as
\begin{equation}
\sigma_{dc} = \frac{\beta^2}{\pi} \Lambda_{xx}(\mathbf{q=0}, \tau = \beta/2),
\label{eq:sigma_dc}
\end{equation}
with
\begin{equation}
\Lambda_{xx}(\mathbf{q}, \tau ) = 
\langle j_{x}(\mathbf{q}, \tau) j_{x}(-\mathbf{q}, 0)  \rangle \,\,.
\end{equation}
$ j_{x}(\mathbf{q}, \tau) $ is the $\mathbf{q}$-$\tau$ dependent current
in $x$ direction, the Fourier transform of $j_{x}(l) = -\mathrm{i} \sum_{l}t_{l
+ \hat{x}, l}^{\phantom{\dagger}} ( c^{\dagger}_{l+\hat{x},\sigma}
c_{l,\sigma}^{\phantom{\dagger}} - c^{\dagger}_{l,\sigma}
c_{l+\hat{x},\sigma}^{\phantom{\dagger}} )$.  The assumptions involved
in the use of Eq.\,\eqref{eq:sigma_dc} to evaluate the conductivity are
discussed in Refs.[\onlinecite{Trivedi96,Denteneer99}] and are tested
there for a variety of situations.

%%%%%%%%%%%%%%%%%%%%%%%%%%%%%%%%%%%%%%%%%%%%%%%%%%%%%%%%%%%%%%%%%%
\section{Single depletion} \label{Single depletion}
%%%%%%%%%%%%%%%%%%%%%%%%%%%%%%%%%%%%%%%%%%%%%%%%%%%%%%%%%%%%%%%%%%

\begin{figure}[t]
\includegraphics[scale=0.32]{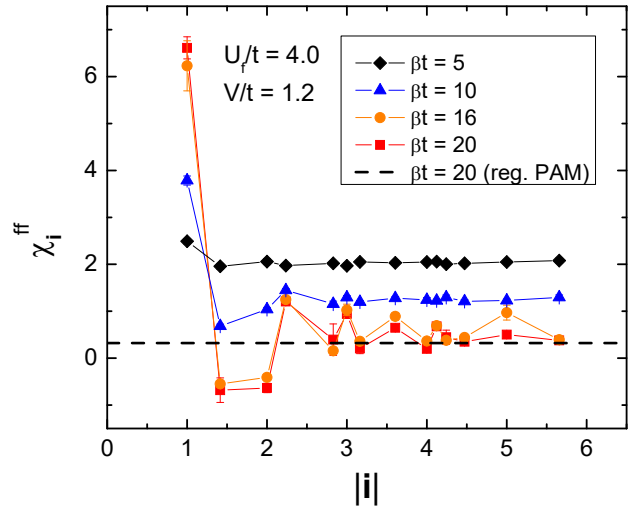}
\caption{(Color online) Local susceptibility of $f$-orbitals as a function
of distance $|\mathbf{i}|$ from the ion defect. The dashed black line is
the value of the magnetic susceptibility in the undepleted
PAM at $\beta t = 20$. Here, and in all subsequent figures, when not
shown, the error bars are smaller than the symbol size.}
\label{Fig:local_chi}
\end{figure}

An interesting step towards understanding $f$-orbital depleted systems
is to consider an isolated impurity. In fact,
there have been a number of recent studies of the alteration of the
magnetic structure around impurities in heavy fermion materials and
their possible description within the PAM\cite{Urbano07,Seo14,dioguardi16,Benali16,Mendes17}.  Hence,
we first discuss how depletion of a single $f$-orbital
affects magnetic and spectral properties, and compare with the uniform
case of the undepleted PAM.  Inclusion of a magnetic defect breaks
translational symmetry.  We therefore generalize Eq.\,\eqref{eq:hom.chi} to
the local magnetic susceptibility at site $\mathbf{i}$ by
\begin{align}
\chi_{\mathbf{i}}^{\alpha\gamma} = \sum_{\mathbf{j}\in \gamma} 
\int_{0}^{\beta} d\tau 
\langle S^{\alpha,\,-}_{\mathbf{i}}(\tau) S^{\gamma\,,+}_{\mathbf{j}}(0) 
\rangle,
\label{eq:localchi}
\end{align}
where the sum runs over all sites ${\bf j}$ with orbital $\gamma$.  The
total susceptibility, Eq.\eqref{eq:hom.chi}, is the sum of these local
susceptibilities.  $\chi^{\alpha\gamma}_{\bf i}$ would be probed
experimentally via nuclear magnetic resonance and, indeed, such site-
and orbital-specific NMR has been used to explore spin and charge
patterns in doped heavy fermion\cite{Seo14,Benali16} and iron-pnictide
superconductors\cite{dioguardi16}.

We analysed a $10\times 10$ lattice, using periodic boundary conditions,
with the depleted site defining the origin of the lattice.   Because
one of the most interesting aspects of site removal is the possibility
of enhancing magnetism\cite{Mendes17}, we fix the hybridization at
$V/t=1.2$ and the repulsive potential as $U_{f}/t=4$, so that we are in
the spin-singlet phase of the undepleted PAM\cite{Vekic95,Hu17}.

\begin{figure}[t]
\includegraphics[scale=0.32]{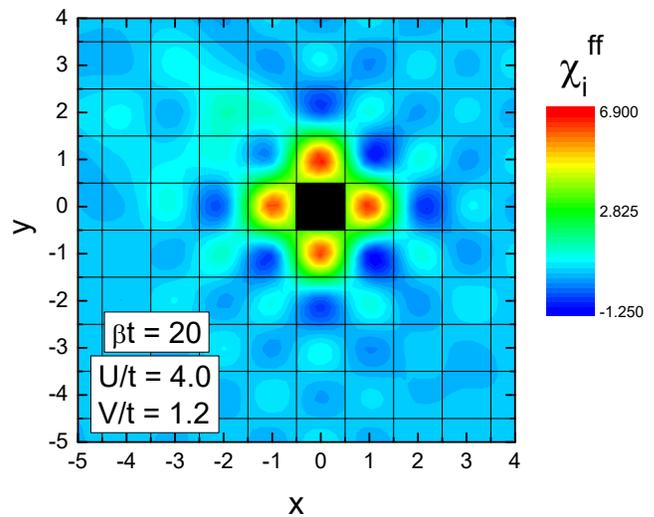}
\caption{(Color online) Contour plot of the $f$-orbital local
susceptibilities for $\beta t = 20$.
The central black square marks the geometrical position of 
the site where the $f$-orbital was depleted.}
\label{Fig:local_chi_contour}
\end{figure}

Figure \ref{Fig:local_chi} presents the behavior of the local
susceptibility, Eq.\,\eqref{eq:localchi}, of the $f$-orbitals as a function of the distance
from the depletion site.  At high temperature, $T/t=0.2$ ($\beta
t=5$, black diamonds), the magnetic response is large and positive,
and almost homogeneous throughout the lattice.  When the temperature is
decreased, $T/t=0.10 - 0.05$, the local susceptibility increases
on nearest neighbor (NN) sites.  This is the opposite of what happens in the
conventional PAM, where, in the singlet phase at $V/t=1.2$, the magnetic
susceptibility goes to zero as a consequence of the spin gap in the
ground state.  Indeed, at $\beta t=20$ (red squares), the NN magnetic
susceptibility is an order of magnitude larger than the undepleted PAM (dashed
black line).  

On the other hand, as $T$ is lowered, the next-nearest neighbors (NNN)
of the defect exhibit a lower, but negative, magnetic response, providing
evidence of antiferromagnetic correlations around the depleted site.  As
the distance from the depleted site grows, $\chi_{\bf i}^{ff}$ decays
with distance, eventually approaching the value for the regular PAM.
An alternate visualization of the
enhancement of antiferromagnetic correlations, is given in
Fig.\,\ref{Fig:local_chi_contour}, a color contour plot of $\chi_{\bf
i}^{ff}$.  The formation of a small antiferromagnetic `cloud' around the
magnetic defect is evident.  When the hybridization is increased (not
shown) the magnetic response for the NN sites remains high, but is
strongly suppressed on sites farther from the impurity.  
The characteristic size of the
`cloud' decreases as one moves deeper into the singlet phase.  The
results of Figs.\,\ref{Fig:local_chi} and \ref{Fig:local_chi_contour} are
consistent with DMRG calculations for a single depletion in the
one-dimensional Kondo Lattice Model (KLM)\citep{CYu96} and with the
behavior of a corresponding model of localized spins; see Ref.\,\onlinecite{Mendes17}.

\begin{figure}[t]
\includegraphics[scale=0.32]{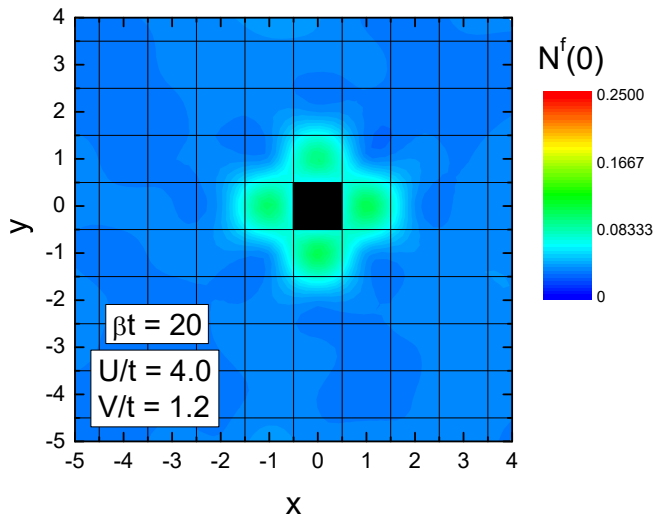}
\caption{(Color online) Contour plot of the local $f$ electron density of 
states, $N^{f}_{\mathbf{i}}(0)$, for
$\beta t = 20$.  The black square corresponds to the impurity location,
as in Fig.~\ref{Fig:local_chi_contour}.}
\label{Fig:local_DOS_Nf_contour}
\end{figure}

As noted earlier, the presence of magnetic clouds around
impurities is a characteristic feature of real materials.  In the heavy
fermion CeCo(In$_{1-x}$Cd$_{x}$)$_{5}$, for example, an
antiferromagnetic region appears around Cd impurities, with a size that
can be tuned with pressure \cite{Urbano07,Seo14}.  In this particular
situation, a more appropriate model than that of Eq.\,\eqref{eq:PAM_hamil}
is one in which the moment on an impurity site has an altered
hybridization $V$ to the conduction electrons\cite{Benali16}.  However,
magnetic domains around sites in which the moment is removed have also
been studied\cite{martin97,azuma97}.

\begin{figure}[t]
\includegraphics[scale=0.32]{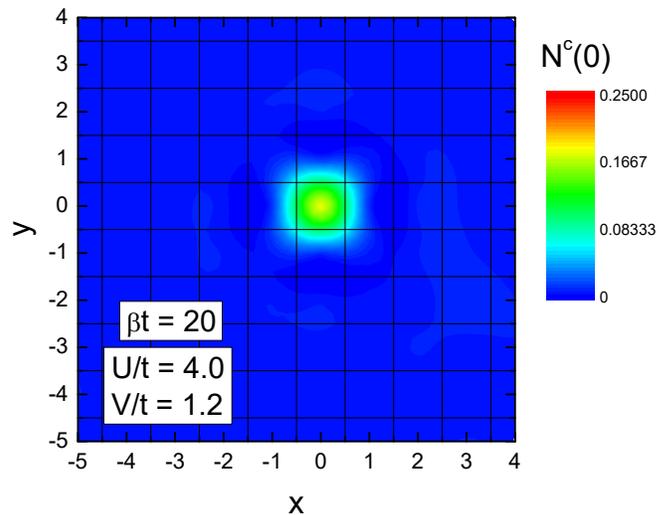}
\caption{(Color online) Contour plot of the conduction electron 
density of states, $N^{c}_{\mathbf{i}}(0)$, for $\beta t = 20$. 
}
\label{Fig:local_DOS_Nc_contour}
\end{figure}

The preceding result is suggestive of the breaking of the local singlet
state, an effect we will see in even more dramatic form when a
collection of $f$ sites is removed.  It is also worth examining the
spectral properties of the system.  We compute the local density of
states (DOS) by analytic continuation of the imaginary-time dependent
Green's function, inverting the integral equation
\begin{align}
G_{\mathbf{i}}(\mathbf{j}=0,\tau) = \int \mathrm{d}\omega\,
N_{\mathbf{i}}(\omega) \,
\frac{e^{-\omega \tau}}{e^{\beta \omega} + 1},
\label{eq:aw}
\end{align}
where $\mathbf{i}$ denotes the site position, while $\mathbf{j}$ is the
displacement between sites where the creation and annihilation operators
of the Green's function are applied.  As discussed in
Ref.\,\onlinecite{Trivedi95}, for low temperatures the DOS at $\omega=0$
can be written as
\begin{align}
N_{\mathbf{i}}(\omega=0) \approx - \beta G_{\mathbf{i}}(\mathbf{j}=0,
\tau=\beta/2) / \pi.
\label{eq:a0}
\end{align}
In Fig.~\ref{Fig:local_DOS_Nf_contour} we present a contour plot for
$N^{f}_{\mathbf{i}}(0)$ at $\beta t = 20$.  At sites far from the
magnetic defect, the local DOS vanishes, as expected in the spin-singlet
phase.  Near the impurity, there is a large DOS, supporting the picture of
broken singlets around the defect, in accordance with the local magnetic
susceptibility results, discussed above.  As displayed in the contour
plot of Fig.~\ref{Fig:local_DOS_Nc_contour}, precisely at the impurity
site the conduction electron DOS, $N^{c}_{\mathbf{i}}(0)$, is large,
owing to the absence of a partner $f$ electron.

\begin{figure}[t]
\includegraphics[scale=0.32]{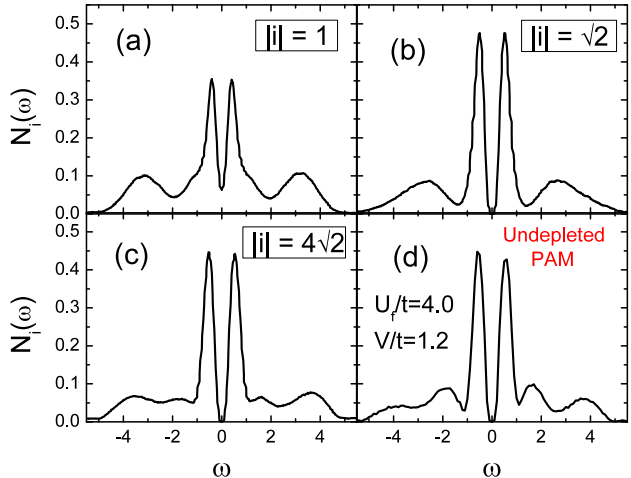}
\caption{DOS of sites at distance (a) $|\mathbf{i}| = 1$, (b) $\sqrt{2}$ and
(c) $4 \sqrt{2}$ from the depleted site, as well as (d) the regular
(undepleted) PAM, both for $\beta t= 20$ and $V/t=1.20$.
At short distances, the singlet gap present in the undepleted PAM
is partially filled in, so that there is nonvanishing $f$
spectral weight at $\omega=0$.}
\label{Fig:local_DOS_Nf}
\end{figure}

To provide an independent check on the validity of Eq.\,\eqref{eq:a0}, we
performed a direct inversion of Eq.\,\eqref{eq:aw}, using the Maximum
Entropy Method\cite{Jarrell96}.  The local DOS is displayed in
Fig.~\ref{Fig:local_DOS_Nf} for sites at (a) $|\mathbf{i}|= 1$
(nearest neighbors), (b) $\sqrt{2}$ (next-nearest neighbors) and (c)
$4\sqrt{2}$ far from the magnetic defect.  As in
Figs.\,\ref{Fig:local_DOS_Nf_contour} and \ref{Fig:local_DOS_Nc_contour},
we fixed $\beta t= 20$.  We also present in Fig.~\ref{Fig:local_DOS_Nf}\,(d) the
DOS of the undepleted PAM, at the same temperature and hybridization.
Notice that for $V/t=1.20$, the undepleted PAM is in the spin-singlet phase
and has a gap in the DOS.  Although the charge gap of the homogeneous
system is recovered at large $|{\bf i}|$, the results suggest that a
single depletion in the spin singlet phase of the PAM creates non-vanishing
spectral weight in the $f$-sites around the defect (and a localized
state in the unpaired $d$ orbital).
Similar analyses within a mean-field approach were performed in
Refs.~\onlinecite{Schlottmann91_1}
and \onlinecite{Schlottmann91_2}, where a gain is also observed in spectral
weight owing to $f$-orbital depletion.

To summarize: our results provide evidence of
enhancement of short range magnetic correlations and local density of
states, which we interpret as arising from the breaking of spin
singlets near the impurity \cite{CYu96,Mendes17}.  We now turn to the
main theme of this manuscript, namely what happens with a regular
collection of depleted sites, and specifically, do the local magnetic
regions coalesce into long range order?  Such problems were first
addressed in Refs.\,\onlinecite{Schlottmann96_2} and
\onlinecite{Fazekas92} for the PAM and KLM, respectively,
within a mean-field approach.
Recent mean-field results for the PAM, presented in Ref.\,\onlinecite{Potthoff14},
also provide evidence of a ferromagnetic ground state.
Similarly, unbiased methods for an analogous spin model, namely the bilayer Heisenberg model
(Ref.\,\onlinecite{Mendes17}),
shows that, indeed, the depletion of spins induces an AFM ground state.
As we shall see in the next Section, magnetic long range order indeed 
occurs, even deep in what was previously the singlet phase of 
the undepleted model (large $d$-$f$ hybridization).
However, as presented below, the system still has a ``memory" of the old critical point.

%%%%%%%%%%%%%%%%%%%%%%%%%%%%%%%%%%%%%%%%%%%%%%%%%%%%%%%%%%%%%%%%%%
\section{Half depletion} \label{Half depletion}
%%%%%%%%%%%%%%%%%%%%%%%%%%%%%%%%%%%%%%%%%%%%%%%%%%%%%%%%%%%%%%%%%%

\begin{figure}[t]
\includegraphics[scale=0.30]{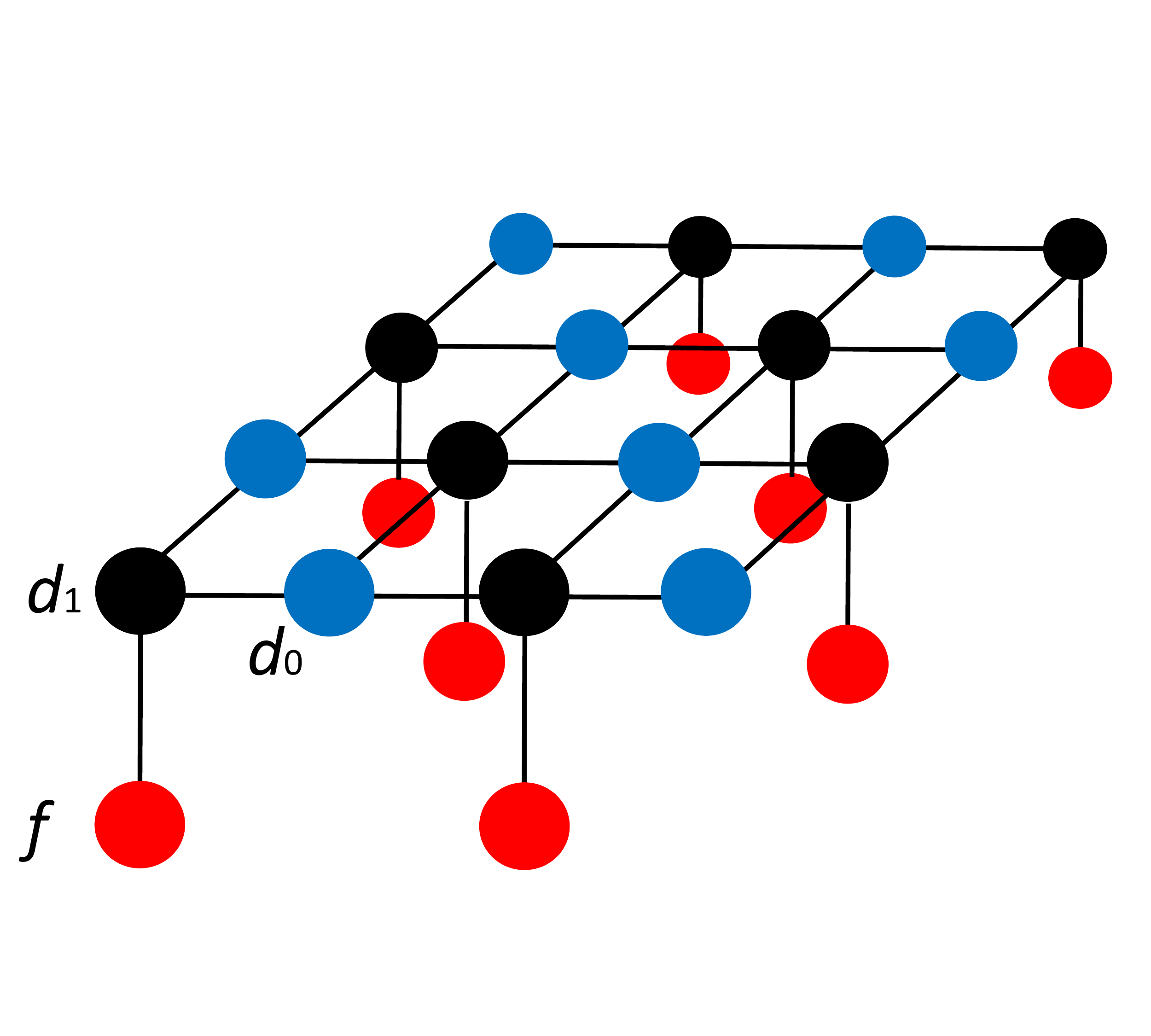}
\caption{(Color online)  The lattice geometry for the
regularly half depleted PAM. The unit cell is composed by the sites
$d_{0}$, $d_{1}$ and $f$.  Here we depict a lattice with
$N_{cells} = 2 \times 2^{2}$.}
\label{Fig:Lattice_half}
\end{figure}

The previous results suggest that if the number of nonmagnetic defects
increases, the magnetic correlations can be enhanced, and thus the ground
state may exhibit magnetic long range order even if the undepleted model
is in the singlet phase.  Here we explore the case of depletion of half
of the $f$-orbitals, in the checkerboard pattern of
Fig.~\ref{Fig:Lattice_half}.  The possibility of a magnetic ground
state in such a geometry is supported by exact results, such as
Tsunetsugu's theorem \cite{Tsunetsugu97} for the KLM, and Lieb's theorem
for the Hubbard model \cite{lieb89,lieb89err,Costa16}.
The former is particularly relevant in this case, owing to the close relationship
between PAM and KLM.
Tsunetsugu showed that the ground state of the KLM on a bipartite lattice and at
half filling has total spin $S=|N_\mathcal{A}-N_\mathcal{B}|/2$, where
$N_\mathcal{A}$ and $N_\mathcal{B}$ are the number of sites in
sublattices $\mathcal{A}$ and $\mathcal{B}$, respectively.
In this theorem, the localized spins are also included when counting the number of
sites in each sublattice, with their labels (i.e. belonging to $\mathcal{A}$
or $\mathcal{B}$) depending on the sign of the Kondo interaction.
There is no assumption of translational symmetry in this
result; missing sites can be randomly located.  Although this theorem
was proved for the KLM, one might expect a similar behavior in the
closely-related PAM.  If so, the total spin of the PAM in the half
depleted lattice of Fig.~\ref{Fig:Lattice_half} should be finite, i.e. a
ferromagnetic ground state.  Our goal here is to confirm this
conjecture within unbiased methods,
and, more importantly, quantify the details of the
individual orbital contributions to the magnetism as a function of the
hybridization between the conduction and localized electrons,
which is beyond the scope of the theorem.

We begin by introducing a notation which simplifies the identification
of the different sites/orbitals.
Since we depleted all $f$-orbitals from one sublattice, then we have two
different types of $d$-orbitals: $d_{0}$ and $d_{1}$.
The former (latter) correspond to the $d$-orbitals without (with)
hybridization with $f$-orbitals, as displayed in Fig.~\ref{Fig:Lattice_half}.
In addition, the unit cell for this geometry is composed by three sites, one of each
type, with unit vectors $\mathbf{a}_{1} = a (1,1) $ and $\mathbf{a}_{2}
= a (1,-1) $. 
Here $a$ is the distance between nearest $d_{0}$ and $d_{1}$
sites.
We will choose $a=0.5$ so that the distance between neighboring
$d_{0}$ (or $d_{1}$) sites is 1.
Finally, for technical reasons we performed our simulations in a $L \times L$ square geometry,
with the primitive cell having twice the size of the unit cell, i.e. 
containing six sites, thus the number of unit cells being $N_{cells}=2 \times L^2$.

\begin{figure}[t]
\includegraphics[scale=0.30]{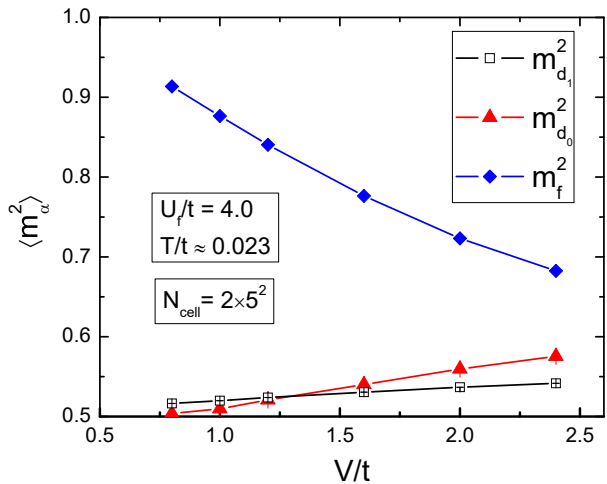}
\caption{(Color online) Local moments on the three types
of sites as functions of hybridization at a low temperature.}
\label{Fig:local_mom}
\end{figure}

We begin by summarizing the noninteracting band structure:  The three-site
unit cell gives rise to three energy bands, the middle of which is
flat (dispersionless), as is also the case for the Lieb lattice.
However, unlike the Lieb lattice for which the flat band touches the
dispersing bands above and below it, here the middle band is
disconnected from the lower and upper bands for any positive value of
$f$-$d$ hybridization $V$.  The system is a band insulator at one-third and
two-thirds filling.  Nevertheless, at half-filling, our focus here, it
is similar to the Lieb case, with a partially filled flat band which,
as we shall see, gives way to bulk ferromagnetism when interactions are
turned on \cite{Vollhardt01}. 

\begin{figure}[t]
\includegraphics[scale=0.40]{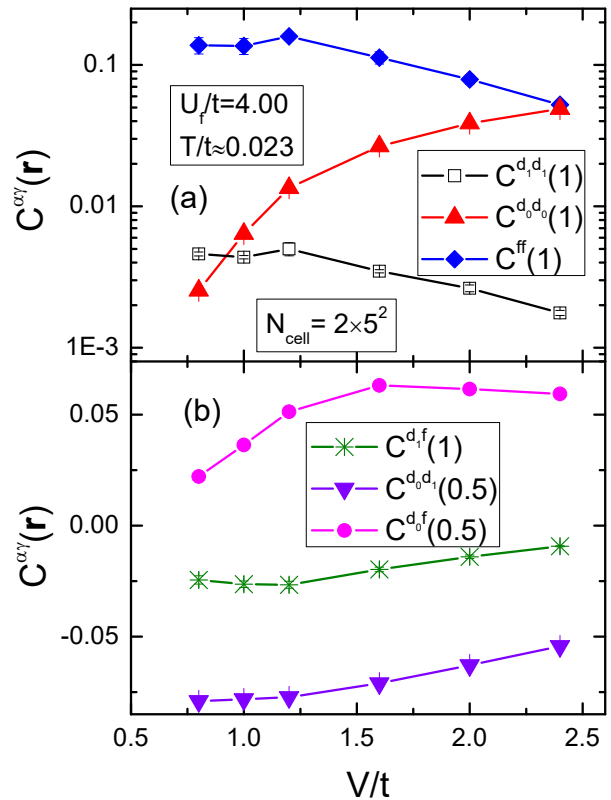}
\caption{(Color online) Spin-spin correlation functions for nearest pair of sites
in the $x$-(or $y$-) direction: (a) $c^{d_{1}d_{1}}(1)$, $c^{d_{0}d_{0}}(1)$,
$c^{ff}(1)$, (b) $c^{d_{0}d_{1}}(0.5)$, $c^{d_{1}f}(1)$ and $c^{d_{0}f}(0.5)$.
Only correlations between pairs of sites on different sublattices,
that is $d_{0}d_{1}$ and $d_{1}f$, are negative (AF).}
\label{Fig:Corr_NN}
\end{figure}

We fix $U_{f}/t=4$ and vary the strength of the $f$-$d$ hybridization $V/t$.
Figure \ref{Fig:local_mom} shows the local moments, $\langle
(\hat{m}^{z}_{\mathbf{i}})^{2} \rangle = \langle (\hat{n}_{\mathbf{i}
\uparrow} - \hat{n}_{\mathbf{i} \downarrow})^{2} \rangle $ on the
different types of sites.  When $V/t \sim 0.8$, inside the AF phase of
the undepleted PAM, $\langle m^{2}_{f} \rangle$ 
(i.e. the local moment of the undeleted $f$-sites)
is large, due to the suppression of the double occupancy $\langle
n_{\uparrow} n_{\downarrow} \rangle $ by $U_f \neq 0$.  By contrast, the
conduction electron moments $\langle m^{2}_{d_{0}} \rangle$ and $\langle
m^{2}_{d_{1}} \rangle$ are close to the half-filled non-interacting value
$\langle m^2 \rangle = \langle n_{\uparrow} + n_{\downarrow} \rangle 
- 2\langle n_{\uparrow} n_{\downarrow} \rangle =\langle n_{\uparrow} +
  n_{\downarrow} \rangle 
- 2\langle n_{\uparrow} \rangle \, \langle n_{\downarrow} \rangle =1/2$.
  As $V$ increases, the hybridization between $d_{1}$ and $f$-sites leads to a
reduction in $\langle m^{2}_{f} \rangle$.  The local moment of the
$d_{0}$-sites increases with $V/t$ much more than that of the $d_{1}$-sites,
which remains roughly constant.  This is somewhat surprising since $V$
connects $d_{1}$-sites directly to $f$-sites, but does not hybridize the $d_{0}$
sites at all,
and, more importantly, $d$-sites have $U=0$.
A similar behavior was recently observed\cite{Mondaini17}
in the Hubbard model on a 2D superlattice with alternating rows of
correlated and uncorrelated sites of different widths.

Further insight into the magnetic properties of the system can be gained by investigating
non-local properties of the real space spin-spin correlation function, Eq.\,\eqref{eq:Cspin}.
We present, in Fig.\,\ref{Fig:Corr_NN}, $C^{\alpha \gamma}(\mathbf{r})$
for nearest pairs of sites along the $x$ (or $y$) direction, as a function
of the hybridization $V/t$.  The sign of the correlations
$C^{\alpha\gamma}(\mathbf{r})$ is always positive for  \{$\alpha \gamma
$\}=\{$d_{0}d_{0}$\}, \{$d_{1}d_{1}$\}, \{$ff$\} and \{$d_{0}f$\}, and negative for \{$d_{0}d_{1}$\}
and \{$d_{1}f$\} even at larger $r$ (not shown).  This is consistent with
Shen's theorem\cite{Shen96} for the KLM, which asserts that, on bipartite lattices,
$C^{\alpha \gamma}(\mathbf{r})$ is always positive for sites on the same
sublattice, and always negative for sites on different sublattices.  

\begin{figure}[t]
\includegraphics[scale=0.31]{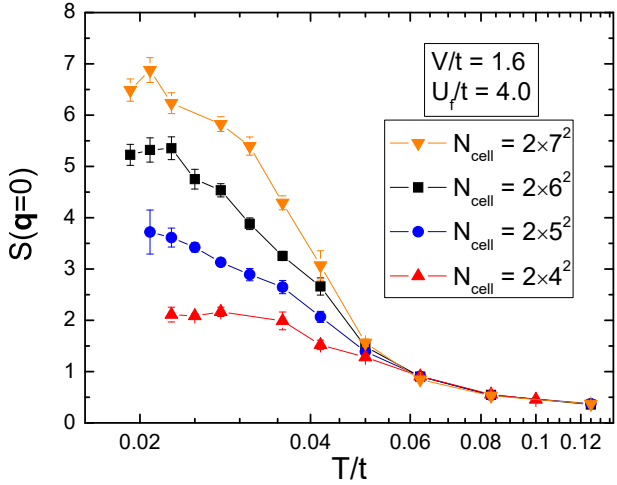}
\caption{(Color online) Homogeneous spin structure factor $S({\bf q}=0)$
versus temperature for $V/t=1.6$ and $U_{f}/t=4.0$.}
\label{Fig:SF}
\end{figure}

In the undepleted case, short range spin correlations decline in
magnitude upon crossing the AF-singlet quantum critical point at $V_c
\sim t$; e.g., see Ref.\,\onlinecite{Hu17}.
Figure \ref{Fig:Corr_NN}, which resolves  the spin
correlations by orbital type, is useful in isolating the origin of the
long range order which we will show to exist, later in this Section.  In
particular, as seen in the figure, some of the short range correlations
grow as $V/t$ increases, contrary to the behavior in the regular PAM.  For
small hybridization, $C^{ff}(1)$ dominates the other correlations; it is
almost two orders of magnitude larger than $C^{d_{0}d_{0}}(1)$, for example.
However, $C^{d_{0}d_{0}}(1)$ is strongly enhanced as $V/t$ increases, while
$C^{ff}(1)$ decreases.  By the time $V/t \sim 2.4$ they are
roughly equal.  Meanwhile, $C^{d_{1}d_{1}}(1)$ remains small at all $V$.  These results
suggest $d_{0}$-sites play an important role in the magnetic
correlations in the ground state.  Fig.\,\ref{Fig:Corr_NN}\, also
indicates that antiferromagnetic correlations are present between
neighboring $d_{0}$ and $d_{1}$-sites [$C^{d_{0}d_{1}}(0.5)$], and $d_{1}$ and $f$-sites
[$C^{d_{1}f}(1)$].  These slowly decrease with $V$.  Finally, $C^{d_{0}f}(0.5)$
exhibits a large and almost constant value, indicating it too 
contributes substantially to ground state  magnetism.

\begin{figure}[t]
\includegraphics[scale=0.30]{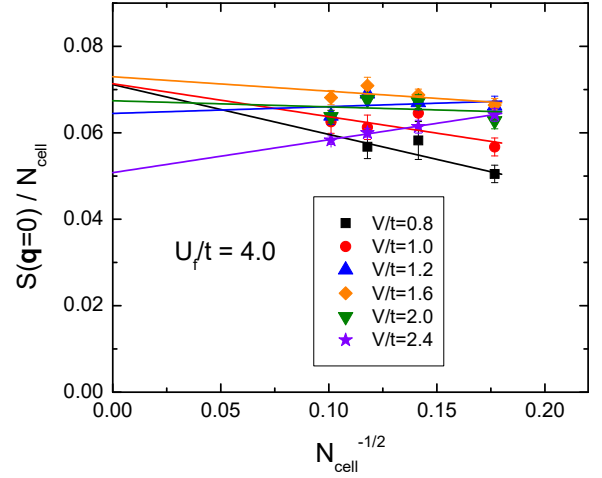}
\caption{(Color online) Finite-size scaling of the total
ferromagnetic structure factor.  A nonzero extrpolation to
$N_{\rm cell}^{-1/2} \rightarrow 0$ indicates the presence
of long range order for all the $V/t$ values shown.
Finite size corrections are largest for small $V/t$ and for large $V/t$;
the effective exchange coupling $J_{\rm eff}$ becomes small in
both limits.  See text.
}
\label{Fig:Scaling}
\end{figure}

According to Eq.\,\eqref{eq:Sq}, the total ferromagnetic spin structure 
factor normalized by the number of sites is defined as
$S(0) = \frac{1}{3} \sum_{\alpha \gamma} S_{\alpha \gamma}$, with
\begin{align}\label{Sq_ab}
S_{\alpha \gamma} = \frac{1}{N_{cells}} \sum_{\mathbf{i}
\mathbf{j}} \langle S^{z(\alpha)}_{\mathbf{i}}
S^{z(\gamma)}_{\mathbf{j}} \rangle \,.
\end{align}
As earlier, $\alpha$ and
$\gamma$ label the sites $d_{0}$, $d_{1}$ and $f$, and the sums over
$\mathbf{i}$ and $\mathbf{j}$ are restricted to their positions.
Fig.\,\ref{Fig:SF} displays the behavior of $S(0)$
at fixed $V/t=1.6$ for
different lattice sizes.  At high temperatures, where spin correlations
are short ranged, this quantity is independent of the size, $N$, of the
system.  However, when the ground state exhibits long range order, the
sum over all sites in Eq.\,\eqref{eq:Sq} becomes dependent on $N$.  More
specifically, in Fig.\,\ref{Fig:SF}, curves with different sizes
separate at $T/t \approx 0.05$, the temperature at
which the correlation length $\xi$ starts to become comparable to the
linear lattice size.  The temperature, which is set by the
effective exchange coupling
$J_{\rm eff}$, where $\xi(T) \sim L$ decreases for larger $V/t$ (not
shown).  As a consequence,
simulations for this parameter regime become very challenging for
DQMC.  The dependence of $J_{\rm eff}$ on $V$ has been studied in
perturbation theory\cite{Potthoff15_1}.

\begin{figure}[t]
\includegraphics[scale=0.45]{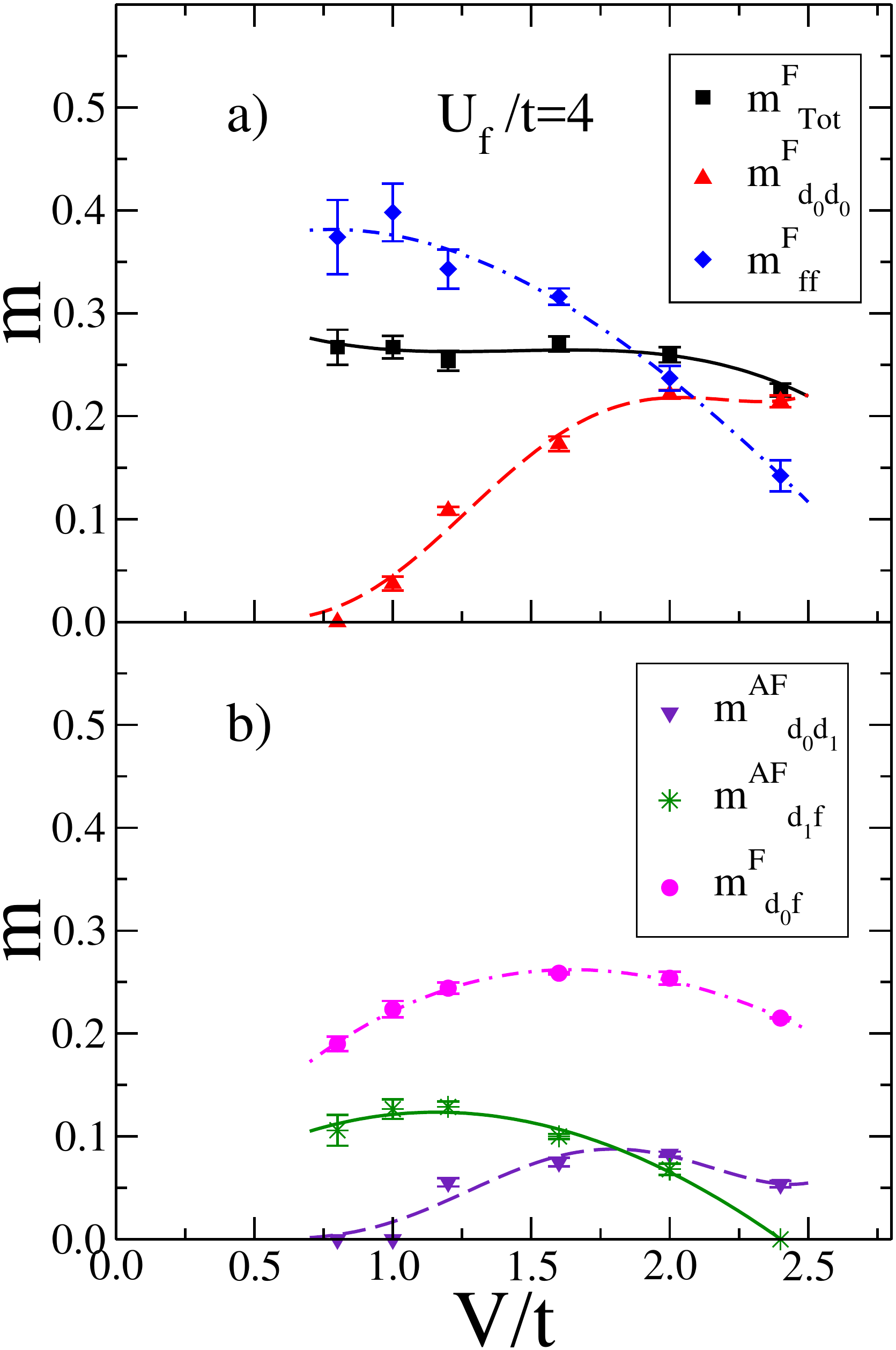}
\caption{(a) The order parameter $m_{tot}^{F}$ and its individual
contributions $m_{d_{0}d_{0}}^{F}$ and $m_{ff}^{F}$ as well as (b)
$m_{d_{0}d_{1}}^{AF}$, $m_{d_{1}f}^{AF}$ and $m_{d_{0}f}^{F}$ as functions of $V/t$.
$m_{d_{1}d_{1}}^{F}$ is zero for all values of $V/t$. The curves are guides
to the eye.}
\label{Fig:phase_diag1}
\end{figure}

The order parameter is obtained by carrying out a finite-size scaling
analysis of the spin structure factor.  The
saturated (large $\beta$) values of $S(0)$, for different lattice sizes
are fit to a linear spin-wave scaling\cite{huse88}, 
\begin{align}
\frac{S(0)}{N_{cell}} =
m^{2}_{tot} + \frac{a}{\sqrt{N_{cell}}},
\end{align}
with $m^{2}_{tot}$ being the
extrapolated global ferromagnetic order parameter; see Fig.~\ref{Fig:Scaling}.
This result confirms the
existence of long range ferromagnetism in the ground state, even for 
$V/t$ more than twice $V_c/t \sim 1$ where the system becomes a spin
liquid in the undepleted case.  
Fig.~\ref{Fig:phase_diag1}\,(a) presents the behavior of
$m_{tot}$ (squares; black solid line) as a function of $V/t$. 
Interestingly, the ferromagnetic order parameter is
almost independent of hybridization $V/t$ over the range shown.  

The analysis of the short range spin correlations for different
orbitals in
Figs.\,\ref{Fig:local_mom} and \ref{Fig:Corr_NN} already provided some
insight into where the magnetism ``lives".  
Then, we proceed by investigating the individual contributions to magnetism,
by means of Eq.\,\eqref{Sq_ab}.
However, as discussed above, the correlation functions of $d_{0}$-$d_{1}$ and
$d_{1}$-$f$ sites are always negative, thus we define 
$S^{AF}_{\alpha \gamma} = -S_{\alpha \gamma}$ for these pairs of sites;
This corresponds to taking their antiferromagnetic contribution.
By the same token, we define $S^{F}_{\alpha \gamma} = S_{\alpha \gamma}$
for those pairs of sites in which their correlation funtions are always ferromagnetic.
As with the global structure factor, we perform a linear
scaling for each channel
\footnote{\label{foot70} The individual contributions $S_{\alpha \gamma}$ 
of channels \{$\alpha \gamma$\}=\{$d_{0}d_{1}$\},
\{$d_{0} f$\} and \{$d_{1} f$\} are divided by two before we perform the scaling.
It assures a site normalized order parameter for all individual contributions.},
i.e. $S^{F (AF)}_{\alpha \gamma}/N_{cells} \to
\big(m^{F (AF)}_{\alpha \gamma}\big)^{2}$, when $1/\sqrt{N_{cells}} \to
0$.  These extrapolated values are displayed in
Fig.~\ref{Fig:phase_diag1}.  We omit $m_{d_{1}d_{1}}^{F}$, which is small
for the entire range of $V/t$ examined.

As shown in Fig.\,\ref{Fig:phase_diag1}\,(a), at small hybridization the
largest contribution to the total magnetism comes from the $f$-sites,
while $m^{F}_{d_{0}d_{0}}$ is negligible.  However, as $V/t$ increases, a
crossover between $f$ and $d_{0}$-site contributions takes place, with the
suppression of the former and the enhancement of the latter,
while the total magnetism is kept constant.
The contributions from different channels are exhibited in Figure \ref{Fig:phase_diag1}\,(b),
with $m^{AF}_{d_{0}f}$ being large over the entire range of $V/t$ we analyzed.
This should not be surprising for small hybridization (i.e. $V/t<1$), since RKKY leads to
long range spin correlations between $f$ and $d$ orbitals.
However, these strong spin correlations even for large $V/t$ ($> 1$) are the key
for supporting the formation of a magnetic ground state.
In this region, attempts to screen the $f$-electrons
reduce their contribution to magnetism, but, owing to the large antiferromagnetic
correlation between $d_{0}$ and $f$-sites, the localized
$d_{0}$-electrons can indirectly interact with each other, leading to 
long range order in their sublattice.
The same assumption can be inferred from $m^{AF}_{d_{0}d_{1}}$.
A similar crossover (from $f$ to $d_{0}$ magnetism) 
is observed in a single spin depleted KLM in a
one-dimensional chain\cite{CYu96}, as well as in higher dimensions
within dynamical mean-field theory (DMFT)\cite{Potthoff14,Potthoff15_2}.  
Unlike DMFT, the DQMC approach
includes nonlocal correlations thus providing additional 
insight into the crossover.

Interestingly, the order parameter on conduction electron sites without 
an $f$ partner, $m^{F}_{d_{0}d_{0}}$, becomes non-negligible at
$V_c/t\approx 1$, the value of the QCP\cite{Vekic95,Hu17}
for the undepleted PAM, for $U_{f}/t=4$. 
In this situation $V_{c}/t$
is the characteristic energy scale to form singlets, whose formation is
prevented on the depleted lattice by the presence of
unpaired $d$-electrons.
One should notice that, not coincidentally, the AF $m^{AF}_{d_{0}d_{1}}$
contribution to magnetism starts being relevant at $V_c/t$ as well.
As discussed above, since $d_{0}$-$d_{0}$ spin correlations are also mediated by
$d_{1}$ sites, one thus expects long range spin correlations in the
$d_{0}$-$d_{1}$ channel in order for $m^{F}_{d_{0}d_{0}}$ be non-negligible.
On the other hand, for large hybridization, namely $V/t \gtrsim 1$, $m^{AF}_{d_{1} f}$ is suppressed, i.e. $d_{1}$-$f$
spin correlations start becoming short ranged, as a symptom of the attempts
to form singlets.
These results strongly suggest that,
despite the fact that magnetism remains present,
there is a ``memory" of the undepleted PAM QCP.
In other words, the $d_{0}$ electrons start being localized and, therefore,
interacting with each other when hybridization is larger than the energy scale
for the formation of singlets in the undepleted PAM.
Thus, we believe that such crossover changes its position according to
the $V_{c}(U_{f})$.
\footnote{Ref.\,\onlinecite{Hu17} presents an accurate determination
of the undepleted PAM QCPs, $V_{c}(U_{f})$.}

%%%%%%%%%%%%%%%%%%%%%%%%%%%%%%%%%%%%%%%%%%%%%%%%%%%%%%%%%%%%%%%%%%
\section{Transport properties} \label{Transport properties}
%%%%%%%%%%%%%%%%%%%%%%%%%%%%%%%%%%%%%%%%%%%%%%%%%%%%%%%%%%%%%%%%%%

\begin{figure}[t]
\includegraphics[scale=0.40]{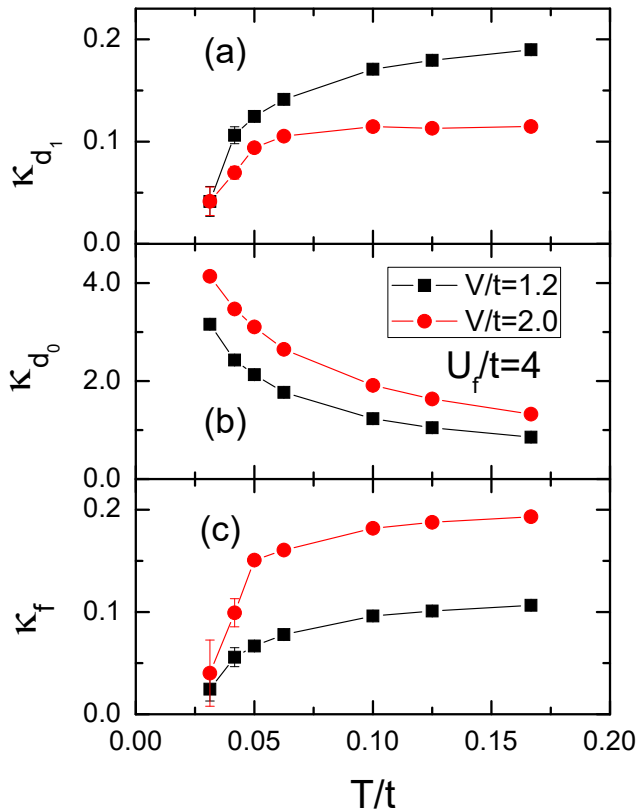}
\caption{
The orbital resolved compressibilities as functions of $T/t$ at
(a) $d_{1}$-, (b) $d_{0}$- and (c) $f$-sites.
On the $f$-sites where $U_f$ is non-zero, and the conduction sites $d_{1}$
to which they are hybridized by $V$, the compressibility vanishes as
$T \rightarrow 0$.  However on the $d_{0}$-sites where only the conduction
orbital remains following depletion, $\kappa$ remains large 
at low $T$.
}
\label{Fig:compress}
\end{figure}

We conclude with a discussion of transport properties.
We first examine the electronic compressibilities
of each individual orbital $\kappa_\alpha$,
which exhibit an interesting behavior (Fig.~\ref{Fig:compress}).  
$\kappa_{d_{1}}$ and $\kappa_f$, the compressibilities on the two sites
connected by $V$, fall as the temperature is lowered
at both $V/t=1.2$ and $V/t=2.0$.
On the other hand, the $d_{0}$-site
compressibility $\kappa_{d_{0}}$ is much larger, and grows as 
$T/t$ decreases.
Such feature of $d_{0}$-sites is deeply connected with the formation of 
localized electronic states on them, as discussed in Sec.\,\ref{Single depletion}.
Thus, one should expect a large density of states at the Fermi level
owing to the absence of local repulsion on $d_{0}$-sites, but with these available
states belonging to them.

\begin{figure}[t!]
\includegraphics[scale=0.30]{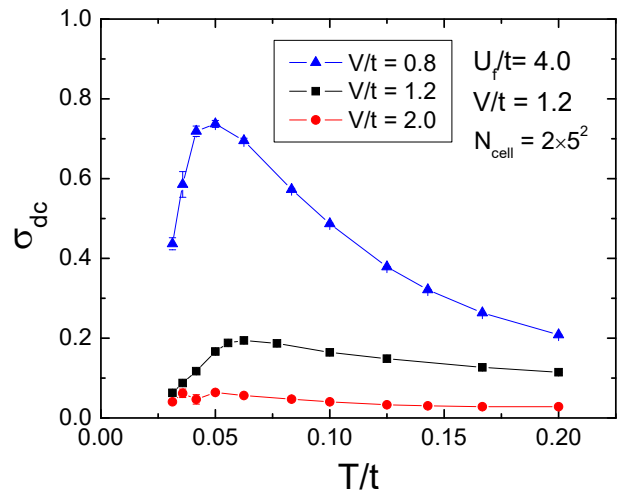}
\caption{
Conductivity $\sigma_{dc}$ as a function of temperature $T$ at
$V/t=0.8$, $1.2$ and $2.0$.  $\sigma_{dc}$ turns sharply downward at the same 
temperature $T \sim 0.05t$ where the compressibility on the $d_{1}$ and $d_{0}$
sites also decreases more rapidly (Fig.~\ref{Fig:compress}) and
the ferromagnetic structure factor signals the beginning
of large spin correlation lengths (Fig.~\ref{Fig:SF}).
}
\label{Fig:conductivity}
\end{figure}
Although the system is compressible, as evidenced by the data
in Fig.\,\ref{Fig:compress}, it is an insulator.
The metallicity of a system should be determined not only by the available states
at Fermi level, but also by its current-current correlation functions and,
ultimately, by its conductivity, Eq.\,\eqref{eq:sigma_dc}.
As displayed in Fig.\,\ref{Fig:conductivity}, the temperature dependence
of the conductivity $\sigma_{dc}$ indicates an insulating behavior.
This is similar to what happens for the single band Hubbard model on a square
lattice, where $\sigma_{dc}$ also vanishes for the entire range of $U/t$, 
as the system crosses over from a Slater to a Mott insulator;
the same also occurs for the entire range of $V/t$ for the undepleted PAM,
whose QCP separates an AF insulator from a singlet phase with
a hybridization gap.  We believe this behavior is generic to the
depleted PAM as well: despite being compressible, the entire half-filled
phase diagram is insulating.

It is worth mentioning that a compressible insulator (or a \textit{gapless}
insulator) was already observed in other systems, such as in one-dimensional
Hubbard superlattices.\cite{Silva02} In this case, a periodic arragement of $L_{1}$
noninteracting and $L_{2}$ interacting sites leads to a compressible insulator
ground state, since one can add charge in the noninteracting sites without energy cost.
By the same token, $U_{d}=0$ in the depleted PAM, in addition to the large local density of
states in $d_{0}$-sites, allows one to accomodate a second electron on them, creating a
compressible (insulator) state.

%%%%%%%%%%%%%%%%%%%%%%%%%%%%%%%%%%%%%%%%%%%%%%%%%%%%%%%%%%%%%%%%%%
\section{Conclusions} \label{Conclusions}
%%%%%%%%%%%%%%%%%%%%%%%%%%%%%%%%%%%%%%%%%%%%%%%%%%%%%%%%%%%%%%%%%%

In this paper we have studied the properties of a two dimensional
periodic Anderson model with (i) a single localized $f$-site depletion
and (ii) one half of the $f$-sites regularly removed.
In the former case, by examining the behavior of the local magnetic
susceptibility, we noticed an enhancement of spin-spin correlations,
with the creation of an antiferromagnetic `cloud' around the defect.
This enhancement occurs because of the break-up of singlets around the
impurity, owing to the exchange interaction between the localized
$f$-electrons with the unpaired $d$-electron.
We have also investigated spectral properties, such as local density of states.
We observed that a single depletion creates a large sprectral weight at the
Fermi level on the unpaired $d$-site, corresponding to a localized state
on it.

The latter case, i.e. the one half $f$-sites depletion, leads to
some unique properties.  First, it has long range ferrimagnetic order,
consistent with Tsunetsugu's theorem\cite{Tsunetsugu97} for the KLM
(and ultimately to Lieb's theorem\cite{lieb89} for the Hubbard Hamiltonian)
concerning the total spin in the ground state of a bipartite lattice with
unequal numbers of sublattice sites.
Analyses of the spin correlations in different channels 
indicate that at small $f$-$d$ hybridization, $V$, 
the magnetic order is dominated by the 
remaining $f$-sites, but at large $V$ there is a crossover:
the magnetic order becomes strongly driven by
those conduction sites which have lost 
their local orbital partners.
It is a remarkable
instance of magnetism from noninteracting orbitals ($U_{d}$=0).
Overall, although the total ferromagnetic order parameter is
surprisingly constant, its individual channel contributions provide
evidence that the crossover between $f$ and $d$ magnetism
occurs at the AF-singlet QCP of the undepleted PAM.

Two additional features stand out in the transport properties.  First,
the system is compressible at half-filling.  This cannot simply be
attributed to the presence of conduction sites at which $U_d=0$, since
those are present in the undepleted PAM, for which $\kappa=0$.  Thus $\kappa \neq 0$
must be attributed to the depletion and, in particular, to the mismatch
of conduction and local spin orbitals which prevents all sites from 
participating in singlet formation.
Related issues have been raised in the reversed situation where the number
of local orbitals exceeds the number of conduction electrons available
for screening\cite{vidhyadhiraja00,meyer00}.
Second, despite this non-vanishing compressibility, the model is
insulating; the conductivity $\sigma_{dc}$ goes to zero as the
temperature is lowered.  

Phases where insulating behavior and nonzero compressibility
are partnered together constitute a prominent feature of
the physics of the {\it boson} Hubbard model\cite{fisher89}.  
There, the introduction 
of disorder results in a new `Bose-glass' phase, which has zero
superfluid density, like the Mott insulator (MI) of the clean model, 
but which is compressible, unlike the MI\cite{scalettar91}.
In the depleted PAM studied here
we have demonstrated a fermionic analog, a phase which is
insulating like the original AF and singlet regimes of the undepleted PAM,
but has nonzero $\kappa$.  This compressible ferrimagnet
originates from depletion rather than from disorder.

%%%%%%%%%%%%%%%%%%%%%%%%%%%%%%%%%%%%%%%%%%%%%%%%%%%%%%%%%%%%%%%%%%
\section*{ACKNOWLEDGMENTS}
%%%%%%%%%%%%%%%%%%%%%%%%%%%%%%%%%%%%%%%%%%%%%%%%%%%%%%%%%%%%%%%%%%

This work was supported by Department
of Energy grant DE-SC0014671 and by the Brazilian Agencies 
CNPq, CAPES, FAPERJ and FAPEPI.

%%%%%%%%%%%%%%%%%%%%%%%%%%%%%%%%%%%%%%%%%%%%%%%%%%%%%%%%%%%%%%%%%%%%%%%%
%%%
%%%%%     BIBLIOGRAPHY
%%%%%%%%%%%%%%%%%%%%%%%%%%%%%%%%%%%%%%%%%%%%%%%%%%%%%%%%%%%%%%%%%%%%%%%%%%%

\bibliography{Pam_depl}

\end{document}